\documentclass[
]{ceurart}

\usepackage{listings}
\usepackage{booktabs}
\usepackage{threeparttable}
\usepackage{graphicx}
\usepackage{subcaption}
\usepackage{float}
\begin{document}

\copyrightyear{2025}
\copyrightclause{Copyright for this paper by its authors.
  Use permitted under Creative Commons License Attribution 4.0
  International (CC BY 4.0).}

\conference{CLEF 2025 Working Notes, September 9 -- 12 September 2025, Madrid, Spain}

\title{Quantum Annealing for Machine Learning: Applications in Feature Selection, Instance Selection, and Clustering}

\title[mode=sub]{Notebook for the QuantumCLEF Lab at CLEF 2025}



\author[1]{Chloe Pomeroy}[%
orcid=0009-0004-2283-7909,
email=cpomeroy6@gatech.edu
]
\fnmark[1]

\author[1]{Aleksandar Pramov}[%
orcid=0009-0005-9049-1337,
email=apramov3@gatech.edu,
]
\fnmark[1]

\author[1]{Karishma Thakrar}[%
orcid=0009-0008-2563-7370,
email=kthakrar3@gatech.edu
]
\fnmark[1]

\author[1]{Lakshmi Yendapalli}[%
orcid=0009-0001-8486-4804,
email=ryendapalli3@gatech.edu
]
\cormark[1]
\fnmark[1]

\address[1]{Georgia Institute of Technology, North Ave NW, Atlanta, GA 30332}
\cortext[1]{Corresponding author.}

 
\fntext[1]{These authors contributed equally.}

\begin{abstract}
This paper explores the applications of quantum annealing (QA) and classical simulated annealing (SA) to a suite of combinatorial optimization problems in machine learning, namely feature selection, instance selection, and clustering. We formulate each task as a Quadratic Unconstrained Binary Optimization (QUBO) problem and implement both quantum and classical solvers to compare their effectiveness. For feature selection, we propose several QUBO configurations that balance feature importance and redundancy, showing that quantum annealing (QA) produces solutions that are computationally more efficient. In instance selection, we propose a few novel heuristics for instance-level importance measures that extend existing methods. For clustering, we embed a classical-to-quantum pipeline, using classical clustering followed by QUBO-based medoid refinement, and demonstrate consistent improvements in cluster compactness and retrieval metrics. Our results suggest that QA can be a competitive and efficient tool for discrete machine learning optimization, even within the constraints of current quantum hardware.
\end{abstract}

\begin{keywords}
  Quantum annealing \sep
  Simulated annealing \sep
  QUBO formulation \sep
  Feature selection \sep
  Instance selection \sep
  Clustering \sep
  DWave Quantum Annealer
\end{keywords}

\maketitle

\section{Introduction}
As machine learning systems are applied to ever-larger datasets, the demands placed on core workflows like feature selection, instance selection, and clustering have grown accordingly. These tasks often involve complex, combinatorial decisions that are challenging to solve efficiently, especially as feature spaces expand into the thousands and datasets span millions of instances. In many cases, classical algorithms struggle to keep up, either becoming computationally prohibitive or falling back on heuristics that don’t guarantee globally optimal solutions.

In response to these challenges, there has been growing interest in leveraging quantum computing paradigms, particularly quantum annealing (QA), for machine learning optimization tasks. By formulating these as Quadratic Unconstrained Binary Optimization (QUBO) problems or Ising models, QA can be applied to select optimal subsets of features or instances, or to identify meaningful clusters. QA offers a fundamentally different mechanism for exploring solution spaces by exploiting quantum tunneling, potentially enabling it to escape local minima more effectively than classical counterparts like simulated annealing (SA). With commercial quantum annealers, such as those provided by D-Wave Systems, now accessible to researchers, it is possible to empirically explore the strengths and limitations of QA in practical machine learning contexts.

The 2025 edition of the Quantum CLEF Competition investigates the feasibility of performing traditional machine learning (ML) tasks by using quantum annealers and comparing their performance to classical methods. It features three subtasks, each to be solved with algorithms runs using both Quantum Annealing (QA) and a Simulated Annealing (SA). Task 1 (Feature Selection) involves selecting the smallest set of features that preserves performance for learning-to-rank on benchmark web collections (MQ2007, ISTELLA) and for an item-based k-NN recommender on a private music corpus with 100- and 400-dimensional item-content matrices. Task 2 (Instance Selection) targets cost-effective fine-tuning of an LLM (Llama 3.1) for sentiment classification by reducing training instances from the Vader NYT and Yelp Reviews datasets without degrading F1 score. Finally, Task 3 (Clustering) requires generating centroid embeddings for the ANTIQUE question-answer corpus, evaluated using the Davies–Bouldin index and query-time nDCG@10 to assess how clustering can accelerate downstream information retrieval tasks ~\cite{qclef2025ceur, qclef2025lncs}.  
 
In this work, we investigate the use of quantum annealing for the three aforementioned core machine learning tasks. We formulate each task as a QUBO problem, suitable for execution on D-Wave’s Advantage\_System quantum annealer. To assess the comparative performance of QA, we also implement classical simulated annealing (SA) using D-Wave’s classical solvers. While quantum annealing remains in its early stages and current hardware imposes certain constraints (e.g., limited qubit connectivity, noise, problem size), our findings show that QA produces competitive solutions and serves as a promising component in hybrid ML pipelines. This work contributes to the growing body of research on the practical viability of quantum optimization for real-world machine learning challenges. All code used in this study is available at the respective GitHub repository for each of the three application areas explored in this work: Feature Selection (\url{https://github.com/dsgt-arc/qclef-2025-feature}), Instance Selection (\url{https://github.com/dsgt-arc/qclef-2025-instance}), and Clustering (\url{https://github.com/dsgt-arc/qclef-2025-clustering}).

\section{Related Work}


Quantum annealing (QA) has emerged as a promising approach for solving combinatorial optimization problems by leveraging quantum fluctuations to escape local minima~\cite{farhi2000quantum}. Unlike gate-based quantum computing, QA is designed to find low-energy solutions to problems expressed as Ising models or, equivalently, as Quadratic Unconstrained Binary Optimization (QUBO) problems. This paradigm has been realized in practical hardware via systems like the D-Wave Advantage, which uses thousands of superconducting qubits~\cite{boothby2020next,yulianti2023systematic}.

Formulating optimization problems as QUBOs is central to harnessing QA effectively. A comprehensive mapping of classical NP-hard problems to QUBO and Ising forms, demonstrating the model's flexibility across domains including graph theory, scheduling, and statistical inference was shown in \cite{lucas2014ising}. Subsequent research extends this work to machine learning, showing how QUBOs can directly encode loss functions and regularization terms for training models \cite{date2023qubo}.

In the context of feature selection (Task 1) and related ML tasks, several studies have explored QA methods using both filter and wrapper approaches.  A QUBO framework that encodes feature importance and redundancy was introduced by \cite{muecke2023feature}, influencing later work by \cite{pranjic2023quantum} and \cite{nembrini2023recommender}, who adapted this for recommender systems. Systematic studies and feature selection techniques, expanding on hybrid solver architectures were done by \cite{borle2023feature} and \cite{yulianti2023systematic}. Repository-scale applications of QA have also emerged: \cite{pranjic2023quantum} performed quantum-annealing-based feature selection in a diverse set of classical supervised learning tasks. 
Feature selection was adapted with QA for recommending content in sparse scenarios, addressing real-world scalability in \cite{nembrini2023recommender}. Their demonstration of combining relevance and redundancy within the QUBO matrix for domain-specific datasets closely aligns with our methodology.

In the context of instance selection (Task 2), \cite{pasin2024quantum} introduced the first Quantum Annealing approach for instance selection problem and indeed proposes the first QUBO formulation for it. It is a straightforward application of cosine similarity between document embeddings and a size constraint encoded into the objective. While their formulation is straightforward, it laid the foundation for subsequent refinements. In parallel, approaches like E2SC\cite{cunha2023effective} and influence-function-based methods~\cite{koh2017understanding, molnar2025, joaquin2024in2core} offer algorithms for instance selection in a classical computing paradigm.
 
With regards to Clustering (Task 3), \cite{bauckhage2019qubo} introduced one of the earliest QUBO formulations for the k-Medoids clustering problem, proposing a binary optimization objective that selects \textit{k} representative medoids from a dataset without requiring explicit cluster assignments. Their formulation directly inspires our refinement stage, as we adopt their objective structure and constraint encoding to enforce fixed-size cluster selection using quantum annealing. Unlike their purely theoretical framing, however, we embed this QUBO formulation into a full pipeline that combines classical pre-clustering with quantum refinement, tailored to work within real-world hardware limits. Building on this, \cite{AlvarezGiron2024qCLEF} applied QUBO-based k-Medoids clustering in a document retrieval context for QuantumCLEF 2024. They implemented a hierarchical method that uses simulated annealing and classical clustering for dimensionality reduction before quantum refinement. Their work demonstrates the promise of combining classical preprocessing with quantum optimization for large-scale embeddings, a structure we also adopt. However, our approach differs by systematically comparing multiple classical clustering methods (e.g., k-Medoids, HDBSCAN, GMM) and integrating a principled formulation of the fixed-\textit{k} constraint using \texttt{dimod.generators.combinations}, enabling more consistent enforcement during sampling. 

QA-ST was proposed by \cite{kurihara2009quantum}, a quantum annealing-based clustering algorithm that extends simulated annealing using a quantum effect to explore multiple suboptimal solutions. Their results show that quantum annealing can outperform simulated annealing (SA) in exploring global optima across datasets such as MNIST and Reuters. While their work focuses on probabilistic exploration within the clustering assignment space, ours emphasizes post-clustering refinement—using quantum annealing to select diverse, high-quality medoids from a pre-clustered pool under strict constraints, which is critical in information retrieval contexts. A novel perspective is contributed by \cite{zaech2023probabilistic},  leveraging all samples returned from a quantum annealer to build calibrated posterior distributions over balanced k-means clusterings. Their probabilistic approach enables uncertainty quantification and ambiguity detection. In contrast, our work prioritizes determinism and fixed-\textit{k} control, optimizing medoid selection to support retrieval performance rather than exploring ensemble uncertainty. A hybrid clustering method is introduced by \cite{matsumoto2022distance}, combining quantum-inspired optimization with classical updates to handle imbalanced data. Their simulated bifurcation method offers fast discrete optimization with high-quality results, yet focuses on cluster balance in traditional assignments. Our pipeline, by contrast, is structured for downstream document retrieval and focuses on interpretability, medoid diversity, and robust fixed-\textit{k} constraints.

In summary, prior research lays important groundwork for QUBO-based clustering and hybrid quantum-classical approaches. Our contribution builds directly on these insights, but advances them through (1) a principled, modular pipeline for real-world document clustering; (2) comparative evaluation of multiple classical clustering strategies upstream of quantum refinement; and (3) robust enforcement of exact medoid count using optimized QUBO constraint encodings. Together, these additions bridge the gap between theoretical clustering formulations and practical, retrieval-oriented quantum applications.

\section{Methodology}

\subsection{QUBO Formulation for Quantum Annealing}
Quantum annealing is a computational process that uses quantum mechanics to find the best solution to complex optimization problems. It relies on the adiabatic theorem of quantum mechanics, which states that a quantum system initially in the ground state of a known, simple Hamiltonian will remain in the ground state if the system evolves slowly enough and the Hamiltonian is changed gradually \cite{farhi2000quantum}. In QA, this principle is used to guide the system from an initial Hamiltonian with a known ground state to a final Hamiltonian that encodes the objective function of an optimization problem. If the evolution follows the conditions of the quantum adiabatic theorem, the system is expected to remain in its ground state, thereby yielding the optimal solution.

To apply quantum annealing to a problem, it must first be formulated as a Quadratic Unconstrained Binary Optimization (QUBO) problem~\cite{lucas2014ising}. A QUBO is defined as:

\begin{equation}
f(\mathbf{x}) = \mathbf{x}^T \mathbf{Q} \mathbf{x}
\end{equation} \label{eq:qubo_general}

\begin{equation}
f(\mathbf{x}) = \sum_{i} Q_{ii} x_i + \sum_{i < j} Q_{ij} x_i x_j
\end{equation}

where $\mathbf{x} \in \{0,1\}^n$ is a binary vector encoding decisions (e.g., feature, instance, or medoid selection), and $\mathbf{Q}$ is an $n \times n$ matrix representing the cost or similarity structure among variables. Q is the QUBO matrix whose diagonal and off-diagonal entries encode the linear weights and pairwise interactions, respectively. The entries of the QUBO matrix \( Q \) can be interpreted in terms of their role in the objective function. The diagonal terms \( Q_{ii} \) represent the linear coefficients associated with individual binary variables \( x_i \), and they determine how much each variable contributes to the total cost when it is set to 1. The off-diagonal terms \( Q_{ij} \) for \( i \ne j \) capture the pairwise interactions between variables \( x_i \) and \( x_j \). A negative off-diagonal entry encourages both variables to take the same value (e.g., both 1), while a positive value penalizes such configurations, promoting diversity or mutual exclusion. This structure allows QUBO to naturally encode constraints and preferences between variables, making it suitable for representing complex optimization problems like feature redundancy minimization or balanced clustering.
The goal of quantum or classical annealing is to find the binary vector $\mathbf{x}$ that minimizes this objective function f($\mathbf{x}$). This formulation serves as the foundation across all tasks in our pipeline, with task-specific adaptations encoded through the construction of $\mathbf{Q}$. While QUBO problems tend to be "unconstrained", we can add a penalty term to the QUBO formulation that allows the problem to have a soft constraint. 

To solve QUBO problems via quantum annealing, we use the D-Wave Advantage\_System4.1 quantum processor. This device consists of 5,760 superconducting qubits laid out in a Pegasus P16 topology, which offers enhanced connectivity and embedding flexibility compared to earlier architectures like Chimera \cite{boothby2020next}. The QUBO problems are submitted through D-Wave’s Ocean SDK\cite{ocean}, which handles the necessary problem embedding, chain construction, and solver parameter configuration. The access to D-Wave's quantum annealers was provided to us by the qCLEF organizers through a specialized infrastructure. 
For comparison, we also evaluate simulated annealing (SA) using D-Wave’s classical solver under similar settings. By running both solvers across the same QUBO formulations, we explore the effectiveness, quality, and consistency of quantum annealing versus classical methods in solving ML-driven optimization problems. The specific formulations for each task are discussed in the following sections.
\begin{figure}[H]
  \centering
  \includegraphics[width=0.4\textwidth]{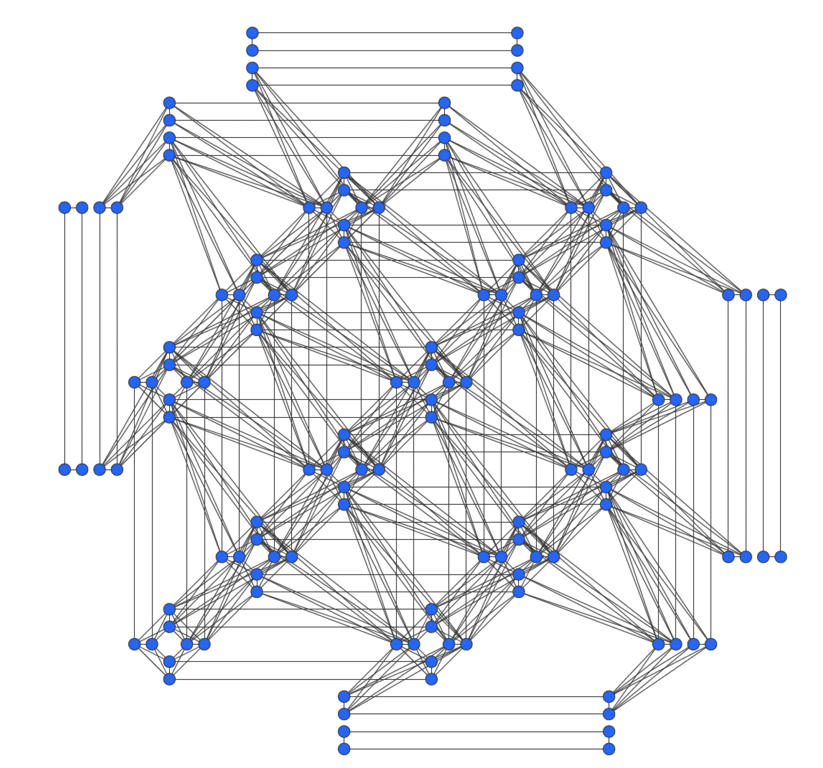}
  \caption{The structure of a Pegasus topology, which defines the connection pattern between qubits in a D-Wave Advantage quantum chip. Blue dots are nodes (qubits) and connecting edges are couplers\cite{pegasus_topology}}
  \label{fig:pegasus}
\end{figure}

\subsection{Task 1: Feature Selection}

Feature selection is a fundamental preprocessing step in many supervised learning pipelines. The goal is to identify a subset of informative, non-redundant features that improve model generalization and reduce overfitting. We formulate feature selection as a combinatorial optimization problem suitable for quantum annealing by leveraging the framework proposed by \cite{muecke2023feature}. Their approach encodes a balance of \emph{feature importance} and \emph{redundancy} directly into a QUBO matrix, making it amenable to solvers like D-Wave.

In our formulation, the \emph{QUBO matrix} \( Q \) is constructed such that:
\begin{itemize}
    \item The \textbf{diagonal entries} \( Q_{ii} \) represent \textbf{importance scores} of individual features.
    \item The \textbf{off-diagonal entries} \( Q_{ij} \) encode \textbf{redundancy} between feature pairs.
    \item A \textbf{penalty term} is included to enforce sparsity and encourage the selection of exactly \( k \) features. This is formulated as a quadratic penalty on the number of selected features, e.g., \( \lambda \left( \sum_i x_i - k \right)^2 \). The penalty term also allows us to explicitly control the number of features selected, tuning \( k \) based on performance.
\end{itemize}

This formulation incentivizes selecting features that are individually relevant while penalizing redundancy and constraining the number of selected features via the penalty term.

\subsection*{Importance and Redundancy Measures}

For the MQ2007 dataset, we evaluated multiple configurations of \( Q \), combining the following measures:

\paragraph{Importance measures (used for \( Q_{ii} \)):}
\begin{itemize}
    \item \textbf{Mutual Information (MI)} between feature \( X \) and target label \( Y \)~\cite{cover2006elements}:
    \[
    \text{MI}(X; Y) = \sum_{x \in X} \sum_{y \in Y} p(x, y) \log \left( \frac{p(x, y)}{p(x)p(y)} \right)
    \]
    \item \textbf{Permutation Feature Importance (PFI)}, defined as the change in model error after permuting feature \( X \)~\cite{breiman2001random}:
    \[
    \text{PFI}(X) = \mathbb{E}[\text{Error}_{\text{perm}(X)} - \text{Error}_{\text{original}}]
    \]
\end{itemize}

\paragraph{Redundancy measures (used for \( Q_{ij} \), \( i \ne j \)):}
\begin{itemize}
    \item \textbf{Conditional Mutual Information (CMI)} between \( X \) and \( Y \) given \( Z \), estimated between pairs of features conditioned on the target~\cite{cover2006elements}:
    \[
    \text{CMI}(X; Y \mid Z) = \sum_{x, y, z} p(x, y, z) \log \left( \frac{p(x, y \mid z)}{p(x \mid z)p(y \mid z)} \right)
    \]
    \item \textbf{Conditional Permutation Feature Importance (CPFI)}, which measures the importance of features \( X \) and \(Y \) when used together~\cite{debeer2020conditional}:
    \[
    \text{CPFI}(X, Y) = \mathbb{E}[\text{Error}_{\text{perm}(X, Y)} - \text{Error}_{\text{original}}]
    \]
\end{itemize}

We experimented with several combinations (e.g., MI+CMI, PFI+CPFI) to populate \( Q \), and selected the combination yielding the best classification accuracy on validation data.

\subsection*{Large-Scale Adaptation for the Istella Dataset}

For the \emph{Istella dataset}, which contains significantly more features, we limited our analysis to the MI+CMI combination due to computational constraints. Notably:

\begin{itemize}
    \item Computing CMI for all feature pairs is computationally expensive. To scale this, we used Python’s \texttt{multiprocessing.Pool()} to parallelize the computation, reducing runtime considerably.
    \item The resulting QUBO matrix was too large to be embedded directly onto the D-Wave Advantage system. To address this, we used the \texttt{LeapHybridSampler()}, which combines classical and quantum resources to solve large QUBOs that exceed qubit count or connectivity limitations.
    
\end{itemize}

This hybrid strategy allowed us to evaluate the viability of QUBO-based feature selection even on larger, real-world datasets.

\subsection{Task 2: Instance Selection}
The second application deals with instance selection, selecting a subset of instances (i.e. a coreset \cite{joaquin2024in2core}) of document embeddings with the goal of fine-tuning an LLM on that selected subset, in a subsequent step. Here we only address the general instance selection challenge as the fine-tuning itself was outside of the scope of the competition. As with all other QA problems, instance selection has to be transformed into a QUBO problem first (possibly by incorporating the constraints in the target function) as per equation \ref{eq:qubo_general}. To that end, we used the backbone of the \emph{bcos} algorithm considered in \cite{pasin2024quantum} which constructs the diagonal and off-diagonal elements of the Q-matrix. Another aspect that we took from \cite{pasin2024quantum} relates to handling of the problem size on the QPU: we batched the dataset in batches of size 80 and processed the data per batch. All the Q matrix entries take that into account - they are calculated on a per-batch basis. 

For the off-diagonal entries $Q_{i,j}$, the \emph{bcos} algorithm it considers two cases between each two embeddings for each document pair $(doc_{i}, doc_{j})$: 
 $-cos(doc_{j}, doc_{i})$ if $doc_{j}$ and $doc_{i}$ have different labels; $cos(doc_{j}, doc_{i})$ if $doc_{j}$ and $doc_{i}$ have the same label. In our work, we kept the off-diagonal entries in same logic as in \emph{bcos} and 
for the \textbf{diagonal} terms $Q_{i, j}$ , where $i = j$,  we investigated the following extensions:
\begin{description}
    \item[\emph{svc-method}] Adding a penalty term in the following way:  First running a simple (in-sample) support-vector-classifier on all documents within a fold, with the document label as a target and the embeddings as features.  Subsequently, by extracting the distance to the fitted support vector of each instance in the fit.  Denoting with $\widehat{d}^{svc}_{i}$ the (estimated) distance to the margin for each instance $i$, we have for each diagonal entry $Q_{i,i} = \frac{1}{ \widehat{d}^{svc}_{i} + 1e-12}$.  Lower distances should get higher weight in the Q-matrix as they are more important for the classification. The entries are subsequently normalized before running the QA step. The hyperparamaters of the support vector classifier here are not that important (after some experimentation we settled an \emph{rbf} kernel with a  gamma parameter $C = 1.0$ for all experiments), as the goal of the distance metric is to establish a relative ranking between the instances. 
    \item[\emph{instance-deletion}] Borrowing motivation from Cook's distance as a measure of influence of a sample point, we ran a simple iterative instance deletion model (logistic regression) measuring the decrease of performance when removing each datapoint \cite[ch. 31]{molnar2025} within a fold. Our goal was to produce a simple heuristic that measures the direct impact of an instance to a classification problem and inspired the choice of the logistic regression as a model that is very fast to compute. The entry for each diagonal element of the Q matrix within a batch $b$ is then simply the value of the influence measure for the effect on the model prediction: 
    \begin{align}
    Q_{i,i} = \frac{1}{n_{b}}\sum_{k=1}^{n_{b}}\left|\widehat{y}_{k} - \widehat{y}_{k}^{(-i)}\right|
    \end{align},
    which is akin to the numerator of Cook's distance, by  changing the functional value from squared distance to absolute distance following [ch. 31]\cite{molnar2025}.
    More complex versions of such measurement instance influence exist and would be subject to further studies, e.g. (\cite{molnar2025}, \cite{pasin2024quantum})  
\end{description}
For our submission, we also included tests with the vanilla \emph{bcos} method. All methods feature an enforced constraint such that the desired level of size reduction is achieved, as in \cite{pasin2024quantum}.

\subsection{Task 3: Clustering}

This research focuses on a document clustering and retrieval pipeline that combines classical machine learning techniques with quantum annealing to address the challenges of working with high-dimensional embedding spaces. The core methodology follows a structured two-stage approach:
\begin{enumerate}
  \item Reduce and summarize the data using classical clustering algorithms (e.g., k-Medoids, HDBSCAN, GMM) to generate candidate medoids.
  \item Apply quantum annealing to refine medoid selection using a constrained QUBO formulation.
\end{enumerate}

The pipeline begins by loading high-dimensional document and query embeddings. To support faster clustering and enable more efficient experimentation, dimensionality reduction using Uniform Manifold Approximation and Projection (UMAP) was explored as an optional preprocessing step. UMAP works by modeling local neighborhood relationships in the high-dimensional space as a graph and then optimizing a low-dimensional representation that preserves both local and global structure. When used, this reduction accelerated the initial clustering process and aided in visualizing the overall document distribution, while reproducibility was ensured through consistent random seeds.

In the first stage, a classical clustering algorithm, selected from k-Medoids, HDBSCAN (Hierarchical Density-Based Spatial Clustering of Applications with Noise), GMM (Gaussian Mixture Model), or a hybrid HDBSCAN-GMM approach, is applied to generate an overcomplete set of candidate medoids. These are representative data points that summarize local structure in the embedding space and serve as a compressed input to the quantum stage. This compression is necessary due to the limited scale of current quantum annealing hardware, which cannot operate over the full embedding space. 

Each clustering algorithm introduces different structural assumptions and was evaluated independently to explore how these influence downstream refinement. K-Medoids was used for its emphasis on compact, interpretable clusters, with automatic selection of $k$ via silhouette and Davies-Bouldin Index optimization. HDBSCAN provided a density-based alternative, able to discover clusters of arbitrary shape and automatically discard low-signal regions as noise. GMM framed clustering probabilistically as such, producing soft memberships that captured overlapping semantic regions in the embedding space. The hybrid HDBSCAN-GMM approach layered these strengths by first isolating dense cores with HDBSCAN and then modeling their uncertainty with GMM. While only one algorithm is used in any given run, this flexibility allowed the pipeline to examine how different clustering assumptions affect the quality and diversity of medoid candidates.

The second stage builds on the general QUBO formulation described in Eq. (1), refining the candidate medoids by solving a constrained optimization problem tailored to clustering. The specific formulation we adopt is based on \cite{bauckhage2019qubo}, which identifies representative medoids without explicitly clustering the data. To compute pairwise dissimilarities between candidate medoids, we use Welsch’s M-estimator, which transforms squared Euclidean distances $D_{ij}$ into robust similarity scores:

\begin{equation}
\Delta_{ij} = 1 - \exp\left(-\frac{1}{2} D_{ij} \right)
\end{equation}

This formulation, also known as the correntropy loss \cite{liu2007correntropy}, emphasizes small distances while suppressing the influence of outliers.

The weighted QUBO objective used for medoid refinement is given by:

\begin{equation}
f(\mathbf{x}) = \mathbf{x}^T \left( \gamma \mathbf{1}\mathbf{1}^T - \frac{\alpha}{2} \Delta \right) \mathbf{x} + \mathbf{x}^T \left( \beta \Delta \mathbf{1} - 2 \gamma k \mathbf{1} \right)
\end{equation}

Here, $\mathbf{x} \in \{0,1\}^n$ indicates medoid selection, $\Delta$ is defined in Eq. (2), $\mathbf{1}$ is the all-ones vector, and $k$ is the desired number of medoids. Following Eq. (3), we set $\alpha = \frac{1}{k}$ and $\beta = \frac{1}{n}$ to normalize contributions from the dispersion and centrality terms, and use $\gamma = 2$ to prioritize the fixed-$k$ constraint. This formulation directly informs our quantum objective matrix $\mathbf{Q}$ and provides principled control over medoid selection behavior.

The QUBO objective (Eq. 5) encodes both pairwise dissimilarities between medoids and a hard constraint enforcing the selection of exactly $k$ clusters, expressed as $\sum_{i=1}^n x_i = k$. This exact constraint is central to the pipeline's design, enabling fixed-$k$ clustering in settings where classical methods often return variable or heuristically chosen cluster counts. We initially experimented with several ways to enforce the fixed-$k$ constraint, including adding a quadratic penalty term $(\sum x_i - k)^2$, post-filtering infeasible samples, and scaling penalty weights. While these worked moderately well with simulated annealing, quantum annealing frequently failed to return exactly $k$ medoids, particularly at small $k$, due to noise and the relatively weak enforcement of linear or diagonal penalties. The quadratic form, while mathematically equivalent, induces pairwise correlations between all variables, creating a steep energy valley that better resists hardware noise and fluctuations. Motivated by this, we shifted to a more principled approach: we first constructed the clustering loss and then applied the fixed-size constraint using \texttt{dimod.generators.combinations}, which implements the same quadratic constraint in a way optimized for quantum hardware \cite{dwave_combinations}. To enforce the fixed-$k$ constraint in practice, we scaled the associated penalty using the maximum energy delta of the clustering term and found that doubling this value consistently stabilized solutions across $k$ and solvers. All quantum and simulated annealing runs used 100 reads per solve. This formulation (Eq. 3) proved to be the most robust across both simulated and quantum annealing settings, offering clean separation between clustering structure and constraint enforcement.

Following refinement, all documents are reassigned to the nearest selected medoid using the original, unreduced embedding space. This separation between reduced-space clustering and full-space evaluation ensures that the final cluster assignments remain faithful to the original data distribution. Cluster quality is measured using the Davies-Bouldin Index (DBI), a metric that balances intra-cluster compactness and inter-cluster separation. To assess retrieval effectiveness, the pipeline matches query embeddings to cluster centroids and ranks documents within each cluster by similarity. Retrieval metrics such as nDCG@10 and relevant document coverage are computed to quantify how well the clusters support downstream information access. 
Overall, this methodology combines the interpretability and scalability of classical clustering with the constraint-enforcing capabilities of quantum optimization. By decoupling the tasks of structural summarization and hard cluster selection, the pipeline makes principled use of quantum resources where they are most effective, optimizing over a reduced, meaningful subset of the data, while retaining the flexibility to experiment with different clustering assumptions upstream.

\section{Results and Discussion}
\subsection{Task 1: Feature Selection}
In this section, we reflect on the results of our experiments across both simulated annealing (SA) and quantum annealing (QA) methods for feature selection. Our primary strategy was to evaluate various combinations of importance and redundancy metrics and to tune the number of selected features, denoted by $k$, to maximize performance on a held-out validation set. Based on this tuning, we selected the best-performing configurations to submit to the qCLEF leaderboard.

For simulated annealing, we explored a range of $k$ values, from 5 to 40 (out of the total 46 features for the MQ2007 data), and analyzed their corresponding performance using both local evaluation (nDCG@10) and leaderboard scores (Table~\ref{tab:sa_results}). Among the different configurations, those involving mutual information and conditional mutual information (MI + CMI) showed strong performance on the validation set. We hypothesize that this may be attributed to the model-agnostic, information-theoretic nature of MI and CMI, which allows for more consistent estimation of feature relevance and redundancy. However, this advantage appears less pronounced on the held-out test set, where configurations based on permutation feature importance (PFI) also performed competitively, particularly when evaluated using LightGBM (LGB) as the underlying model. Notably, LGB-based methods produced the highest nDCG scores in our local experiments. Unfortunately, we were unable to submit LGB-based feature sets to the shared qCLEF evaluation infrastructure due to compatibility issues. The LightGBM package relies on system-level OpenMP support, specifically the \texttt{libgomp.so.1} shared library. This library was not provided in our restricted qClef development environment, leading to runtime import errors and preventing the use of LightGBM. Thus, we were limited to using XGBoost (XGB) for official submissions. This constraint may have impacted the final leaderboard performance of otherwise stronger feature selection combinations.

\begin{table}[H]
\centering
\caption{nDCG@10 scores across different feature selection strategies using Simulated Annealing (SA). Each strategy corresponds to a combination of importance measures (e.g., MI, PFI) used on the diagonal of the QUBO matrix and redundancy measures (e.g., CMI, CPFI) used on the off-diagonal.}
\label{tab:sa_results}
\resizebox{\textwidth}{!}{%
\begin{tabular}{@{}ccccccc@{}}
\toprule
\textbf{$k$} & \textbf{MI + CMI} & \textbf{PFI + CMI (lgb)} & \textbf{PFI + CPFI (lgb)} & \textbf{PFI + CPFI (xgb)} & \textbf{PFI + CMI (xgb)} \\
\midrule
5  & 0.3817 & 0.2002 & 0.1898 & 0.3811 & 0.3671 \\
10 & 0.5232 & 0.6509 & 0.2474 & 0.1931 & 0.4865 \\
15 & 0.5897 (0.4485)$^{\ddagger}$ & 0.3431 & 0.4531 & 0.5215 & 0.1847 \\
20 & 0.4670 & 0.4142 & 0.5419 & 0.5745 (0.4318)$^{\ddagger}$ & 0.3255 \\
25 & 0.5859 (0.4510)$^{\ddagger}$ & 0.5123 & 0.4655 & 0.5396 & 0.5845 (0.4500)$^{\ddagger}$ \\
30 & 0.4003 & 0.4675 & 0.6873 & 0.5646 & 0.5679 (0.4523)$^{\ddagger}$ \\
35 & 0.3970 & 0.6211 & 0.4312 & 0.5398 & 0.5688 \\
40 & 0.3866 & 0.3866 & 0.5781 & --     & --     \\
\bottomrule
\end{tabular}%
}
\vspace{0.5em}
\begin{tablenotes}
\small
\item[] Each cell shows the validation nDCG@10 score. These were calculated on the validation set.
\item[] The baseline nDCG@10 score including all 46 features is 0.4473. 
\item[] $^{\ddagger}$ Configuration submitted to the qCLEF leaderboard. Values in parentheses are the official CLEF leaderboard scores calculated on the held-out test set. 
\item[] MI: Mutual Information; PFI: Permutation Feature Importance; CMI: Conditional Mutual Information; CPFI: Conditional Permutation Feature Importance.
\item[] PFI and CPFI were computed using LightGBM (lgb) and XGBoost (xgb) based importance scores.
\item[] All results shown are from local validation runs. While LightGBM (LGB) models often outperformed XGBoost (XGB) in our local evaluations, we were unable to submit LGB-based models to the shared qCLEF evaluation infrastructure due to compatibility issues. As a result, only XGB-based feature sets were submitted for final leaderboard scoring.
\item[] “--” indicates configurations that were not evaluated due to resource constraints.
\end{tablenotes}
\end{table}

For quantum annealing, despite our interest in conducting experiments for more configurations, we were constrained by limitations in the quantum infrastructure. Specifically, the time and resource availability for the D-Wave quantum annealer limited the breadth of our QA experiments. As a result, we were only able to submit two QA-based runs, both derived from the same codebase and configuration (Table \ref{tab:qa_results}). Interestingly, these two QA submissions resulted in different outcomes: one returned a feature subset of size 13 with an nDCG of 0.4552, while the other selected 15 features and achieved an nDCG of 0.4436. This divergence is notable because the code and QUBO formulation were identical in both cases. We attribute this variance to the inherent randomness and probabilistic nature of the quantum annealing process, where solution quality can fluctuate between runs due to quantum noise, minor differences in embedding, or hardware-level stochasticity.

\begin{table}[H]
\centering
\caption{CLEF leaderboard nDCG@10 scores for Quantum Annealing (QA)-based feature selection. All configurations used the D-Wave Advantage\_system quantum annealer and Mutual Information (MI) as the importance measure on the diagonal.}
\label{tab:qa_results}
\begin{tabular}{@{}ccc@{}}
\toprule
\textbf{Method} & \textbf{$k$} & \textbf{Leaderboard nDCG@10} \\
\midrule
QA (MI-CMI) & 15 & 0.4436 \\
QA (MI-CMI) & 13 & 0.4552 \\
\bottomrule
\end{tabular}
\vspace{0.5em}
\begin{tablenotes}
\small
\item[] These configurations were submitted directly to the CLEF leaderboard without local validation.
\item[] The QA runs were executed using D-Wave’s Advantage\_system with 5760 qubits and Pegasus topology.
\end{tablenotes}
\end{table}

An interesting result is that the QA submission with just 13 features achieved the highest nDCG score among all our submissions, and notably, it also had the fewest selected features among all leaderboard entries. While the top leaderboard entry achieved an nDCG of 0.4580 using 21 features, our QA submission reached a comparable score of 0.4552 with only 13 features. This makes it arguably the most efficient feature subset in terms of predictive performance per feature used.

\begin{figure}[H]
\centering
{\large \textbf{Simulated vs Quantum Annealing Results}}
\vspace{0.3cm}

\begin{subfigure}{0.45\textwidth}
    \centering
    \includegraphics[width=\textwidth]{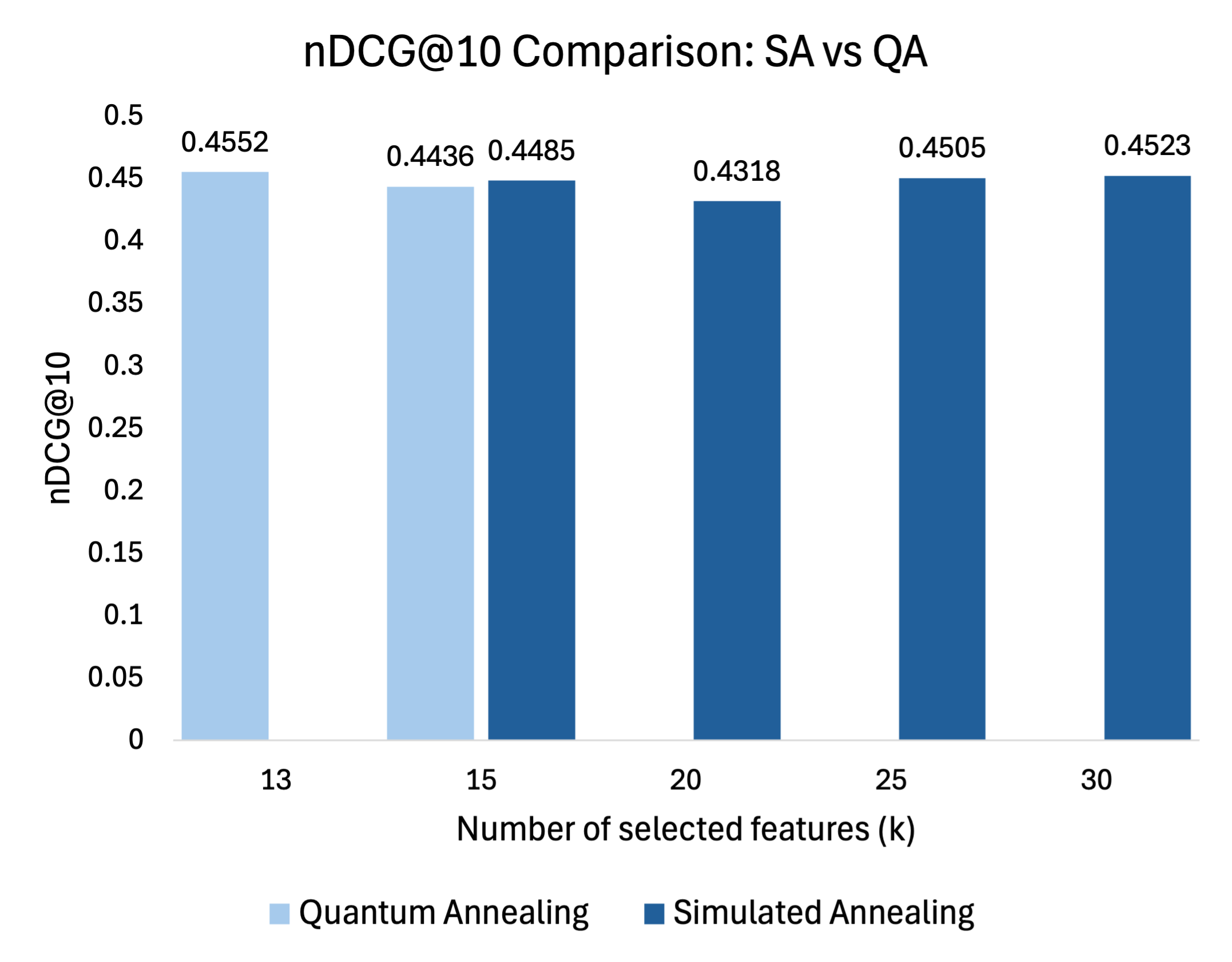}
    \caption{nDCG@10 comparison across SA and QA.}
    \label{fig:task1_ndcg}
\end{subfigure}
\hfill
\begin{subfigure}{0.45\textwidth}
    \centering
    \includegraphics[width=\textwidth]{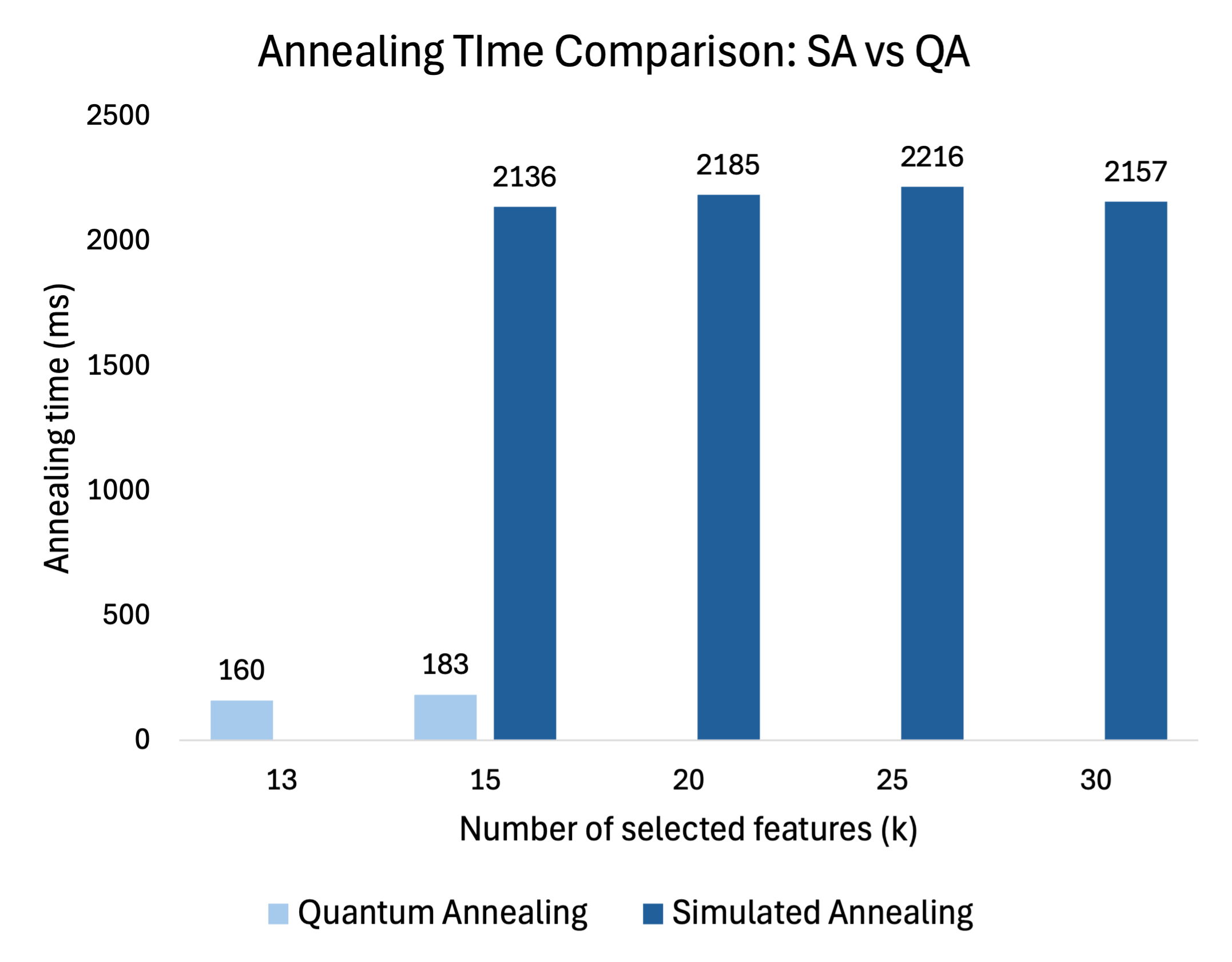}
    \caption{Annealing time (ms) for SA and QA.}
    \label{fig:task1_time}
\end{subfigure}

\caption{Comparison of Simulated Annealing (SA) and Quantum Annealing (QA) on feature selection task. Values in the chart only reflect the submissions made to the qCLEF leaderboard and official results.}
\label{fig:task1_results}
\end{figure}

Furthermore, while SA and QA achieved comparable nDCG scores across the board, the computational effort differed significantly (Figure \ref{fig:task1_results}). Our analysis shows that QA completed the optimization process in approximately one-tenth the time required by SA for similar configurations (see \ref{fig:task1_time}). This suggests that quantum annealing may offer a more efficient route to high-quality solutions, especially in time-sensitive or resource-constrained environments. Overall, these findings underscore the potential of quantum annealing not just as a novelty, but as a competitive alternative to classical metaheuristics like simulated annealing for tasks like feature selection in machine learning pipelines.

\subsection{Task 2: Instance Selection}

The way to properly evaluate an instance selection routine in the context of a QA routine is based on a tripod of criteria, as noted in \cite{pasin2024quantum}: size reduction, performance and total inference time. Naturally, there can be a situation where no method dominates the others in all three aspects. We decided to focus on F1 score, as there was no guideline regarding the reduction size - all of our submissions were at a size of a targeted reduction size of $25\%$ - i.e. 75\% of the instances were targeted to be kept. Table \ref{tab:instance} shows the competition results achieved by our team. We had managed to perform one quantum run which we kept - using the \emph{bcos} method, which does not achieve exact 25\% reduction due the inherent randomness in the quantum annealing procedure, but all the simulated annealing runs are at exacty 25\% reduction. As evident by the standard deviation numbers (in brackets), while we nominally top the leaderboard for a fixed reduction size, the differences to the baseline and to the other teams' submissions are not statistically significant for the \emph{yelp} dataset. For the \emph{vader} dataset all teams perform worse than the baseline, which remains puzzling as our own analysis indicates a much higher performance than indicated by the leaderboard. 

\begin{table}[H]
\centering
\caption{DS@GT and Baseline results with overall rankings (rank is against \emph{all} submissions for each dataset)}
\label{tab:instance}
\begin{tabular}{c l l c c c c}
\hline
\textbf{Rank} & \textbf{Team} & \textbf{Name} & \textbf{F1 Score} & \textbf{Size} & \textbf{Time} & \textbf{Type} \\
\hline
\multicolumn{7}{c}{\textbf{Yelp Dataset}} \\
\hline
1 & DS@GT qClef & Yelp\_SA\_qclef\_bcos\_075            & 99.5(0.2)  & 0.25  & 1548.5(2.8)  & S \\
2 & DS@GT qClef & Yelp\_QA\_qclef\_bcos                 & 99.4(0.2)  & 0.274 & 1500(54.7)   & Q \\
2 & Baseline    & BASELINE\_ALL                        & 99.4(0.1)  & --    & 2027.1(1.1)  & -- \\
3 & DS@GT qClef & Yelp\_SA\_qclef\_it\_del\_075         & 99.3(0.3)  & 0.25  & 1549.2(1.5)  & S \\
3 & DS@GT qClef & Yelp\_SA\_qclef\_svc\_075             & 99.3(0.4)  & 0.25  & 1550.5(2.6)  & S \\
\hline
\multicolumn{7}{c}{\textbf{Vader Dataset}} \\
\hline
1 & Baseline    & BASELINE\_ALL                        & 88.9(0.8)  & --    & 1997.3(5.7)  & -- \\
2 & DS@GT qClef & Vader\_SA\_qclef\_combined\_075       & 65.9(4.7)  & 0.25  & 1529.4(3)    & S \\
3 & DS@GT qClef & Vader\_SA\_qclef\_it\_del\_075        & 65.6(3)    & 0.25  & 1529.5(2.3)  & S \\
4 & DS@GT qClef & Vader\_SA\_qclef\_svc\_075            & 65.4(7.1)  & 0.25  & 1529(2.4)    & S \\
7 & DS@GT qClef & Vader\_QA\_qclef\_bcos                & 62.6(7.5)  & 0.283 & 1493.3(83)   & Q \\
8 & DS@GT qClef & Vader\_SA\_qclef\_bcos\_075           & 62.5(10.4) & 0.25  & 1528.6(2.2)  & S \\
\hline
\end{tabular}
\end{table}

\begin{figure}[H]
  \centering
  \includegraphics[width=0.65\textwidth]{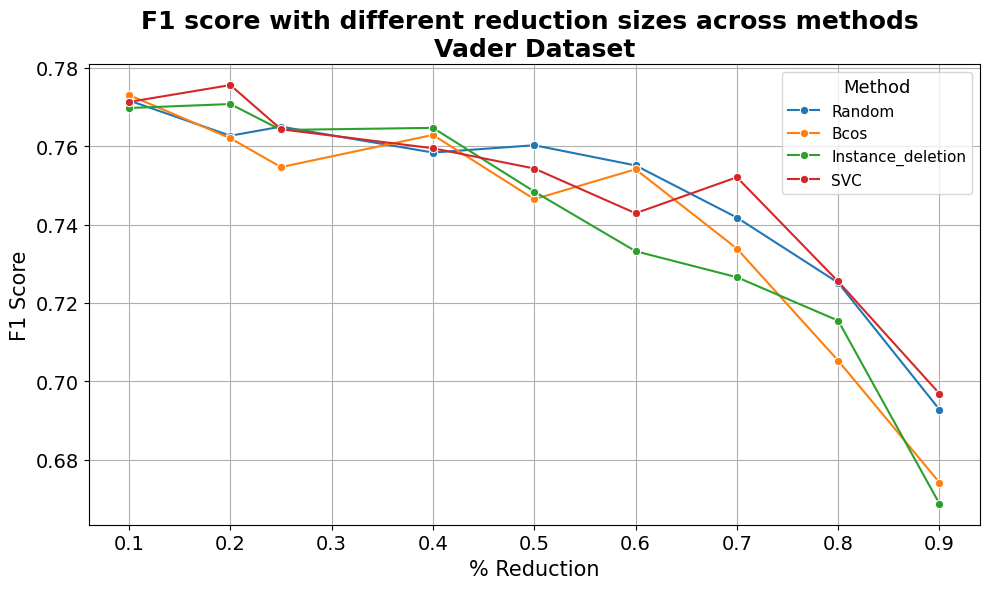}
  \caption{Vader dataset: Average Performance (F1 Score) across test folds of a logistic regression model ran with different instance selection variants and reduction sizes: random sampling, distance to support-vector based method (svc) as well as instance deletion method.}
  \label{fig:evaluation_instance}
\end{figure}

One insight that can be inferred is that the datasets are simply too trivial for this task. \cite{pasin2024quantum} also analyzed these datasets (albeit not in the context of LLM, but of BERT models as a downstream task) and recorded performance of the reduced dataset (by 25\%) at the level of the the full dataset's performance. Another argument that supports this insight can is illustrated by the analysis depicted in figure \ref{fig:evaluation_instance}. In it, we performed (on the test sets across the different folds provided by the organizer) simple logistic regression model as a substitute of the LLM fine-tuning step, which was not accessible to us as competition participants. The average (across test folds) F1 score is shown for all methods, including a simple random sample method, which drops 25\% of the observations within a training fold randomly. They remain fairly stable even for high levels of reduction (10\% to 60\%), as evidenced also by other teams' submissions on the leaderboard who submitted runs with higher reduction level. We also conducted experiments with fine-tuning BERT models and results were comparable - at the 25\% reduction level, there was not a significant difference between a random sampling method and all other methods.

Overall, the (simple) heuristics presented here do not exhibit a significant difference across different reduction levels - neither on the LLM evaluations done on the leaderboard, nor in the simple logistic regression evaluations shown in Figure \ref{fig:evaluation_instance}. Likely, this is due to the size of the dataset and the (low) difficulty of the classification task. A more difficult benchmark dataset could help study these differences in higher detail.

Nonetheless, the SVC-method does show certain promise as being the best performing at higher reduction levels and could be a good starting point for further research. 
\subsection{Task 3: Clustering}
 
\begin{table}[H]
\centering
\caption{Clustering performance across different classical clustering methods and configurations (Runs 1–18), all followed by the same quantum-constrained refinement stage. Baselines are fully classical with no quantum post-processing.}
\label{tab:clustering_results}
\resizebox{\textwidth}{!}{%
\begin{tabular}{@{}cccccccc@{}}
\toprule
\textbf{} & \textbf{Classical Medoid} & \textbf{k} & \textbf{UMAP-} & \textbf{DBI} & \textbf{DBI} & \textbf{nDCG} & \textbf{nDCG} \\
& \textbf{Generation} & & \textbf{Reduced} & & \textbf{(leaderboard)} & & \textbf{(leaderboard)} \\
\midrule
1 & k-Medoids & 10 & N & 7.4776 & 7.4776 & 0.48 & 0.58 \\
2 & k-Medoids$^*$ & 10 & Y & 6.2839 & 4.4706 & 0.50 & 0.0172 \\
3 & k-Medoids & 25 & N & 6.2554 & -- & 0.48 & -- \\
4 & k-Medoids & 25 & Y & 5.1365 & -- & 0.44 & -- \\
5 & k-Medoids & 50 & N & 4.6063 & -- & 0.36 & -- \\
6 & k-Medoids & 50 & Y & 3.7141 & -- & 0.32 & -- \\
7 & HDBSCAN & 10 & N & 3.1862 & -- & 0.52 & -- \\
8 & HDBSCAN & 10 & Y & 6.1444 & -- & 0.56 & -- \\
9 & HDBSCAN & 25 & N & 5.2627 & -- & 0.46 & -- \\
10 & HDBSCAN & 25 & Y & 5.0119 & -- & 0.46 & -- \\
11 & HDBSCAN & 50 & N & 4.3931 & -- & 0.50 & -- \\
12 & HDBSCAN$^*$ & 50 & Y & 3.9469 & 3.4217 & 0.44 & 0.0064 \\
13 & GMM & 10 & Y & 5.9105 & -- & 0.60 & -- \\
14 & GMM & 25 & Y & 5.2283 & -- & 0.36 & -- \\
15 & GMM & 50 & Y & 4.5136 & -- & 0.38 & -- \\
16 & HDBSCAN-GMM & 10 & Y & 6.2197 & -- & 0.48 & -- \\
17 & HDBSCAN-GMM & 25 & Y & 5.1848 & -- & 0.54 & -- \\
18 & HDBSCAN-GMM & 50 & Y & 3.9478 & -- & 0.50 & -- \\
\midrule
-- & Baseline & 10 & -- & -- & 7.9892 & -- & 0.5509 \\
-- & Baseline & 25 & -- & -- & 6.1201 & -- & 0.5284 \\
-- & Baseline & 50 & -- & -- & 5.3679 & -- & 0.4656 \\
\bottomrule
\end{tabular}%
}
\begin{tablenotes}
\small
\item[] DBI: Davies-Bouldin Index (lower is better); nDCG: Normalized Discounted Cumulative Gain (higher is better).
\item[] Internal scores are computed on the training set; leaderboard scores reflect performance on the held-out test set.
\item[] Experiments 1, 2, and 12 reflect submitted results with experiment 1 achieving top score for the task.
\item[] *Dimensionality-reduced centroids were included in the final submission, leading to evaluation errors.
\end{tablenotes}
\end{table}


We evaluated a range of clustering configurations to explore how different classical methods and quantum refinement strategies impact retrieval effectiveness. Table \ref{tab:clustering_results} summarizes the results from both submitted and exploratory experiments. Among the submitted runs, Experiment 1 (k-Medoids, $k=10$, no UMAP) achieved the highest performance, with a leaderboard nDCG of 0.58 and an internal validation score of 0.48, outperforming all baselines. This strong result can be attributed to the simplicity and structure-preserving nature of the two-step k-Medoids pipeline, which maintained the original semantic geometry of the embedding space and yielded consistently strong retrieval performance.

\begin{figure}[H]
\centering
{\large \textbf{K-medoids Clustering Results (Experiment 1)}}
\vspace{0.3cm}

\begin{subfigure}{0.48\textwidth}
    \centering
    \includegraphics[width=\textwidth]{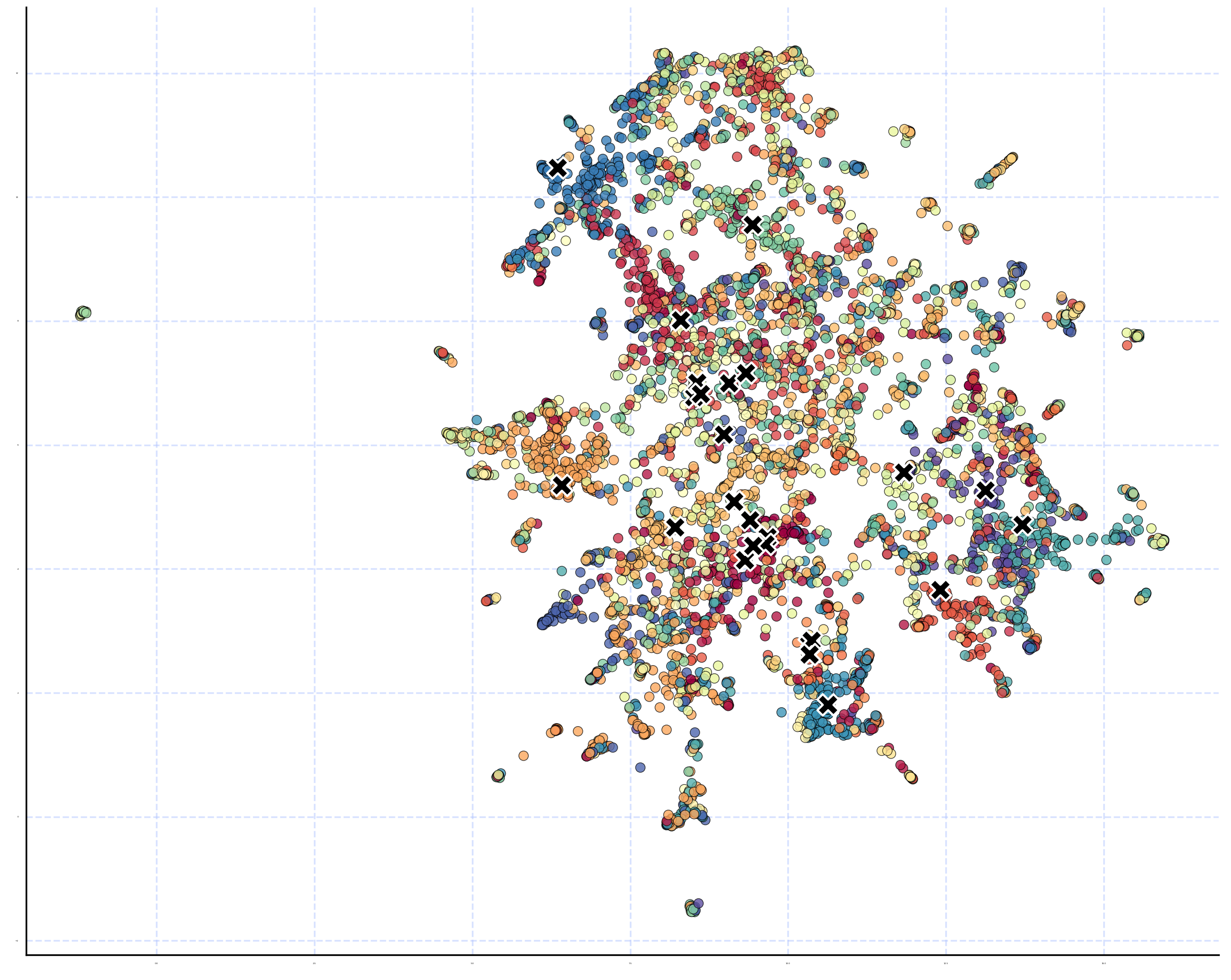}
    \caption{Initial clustering}
    \label{fig:run1_initial}
\end{subfigure}
\hfill
\begin{subfigure}{0.48\textwidth}
    \centering
    \includegraphics[width=\textwidth]{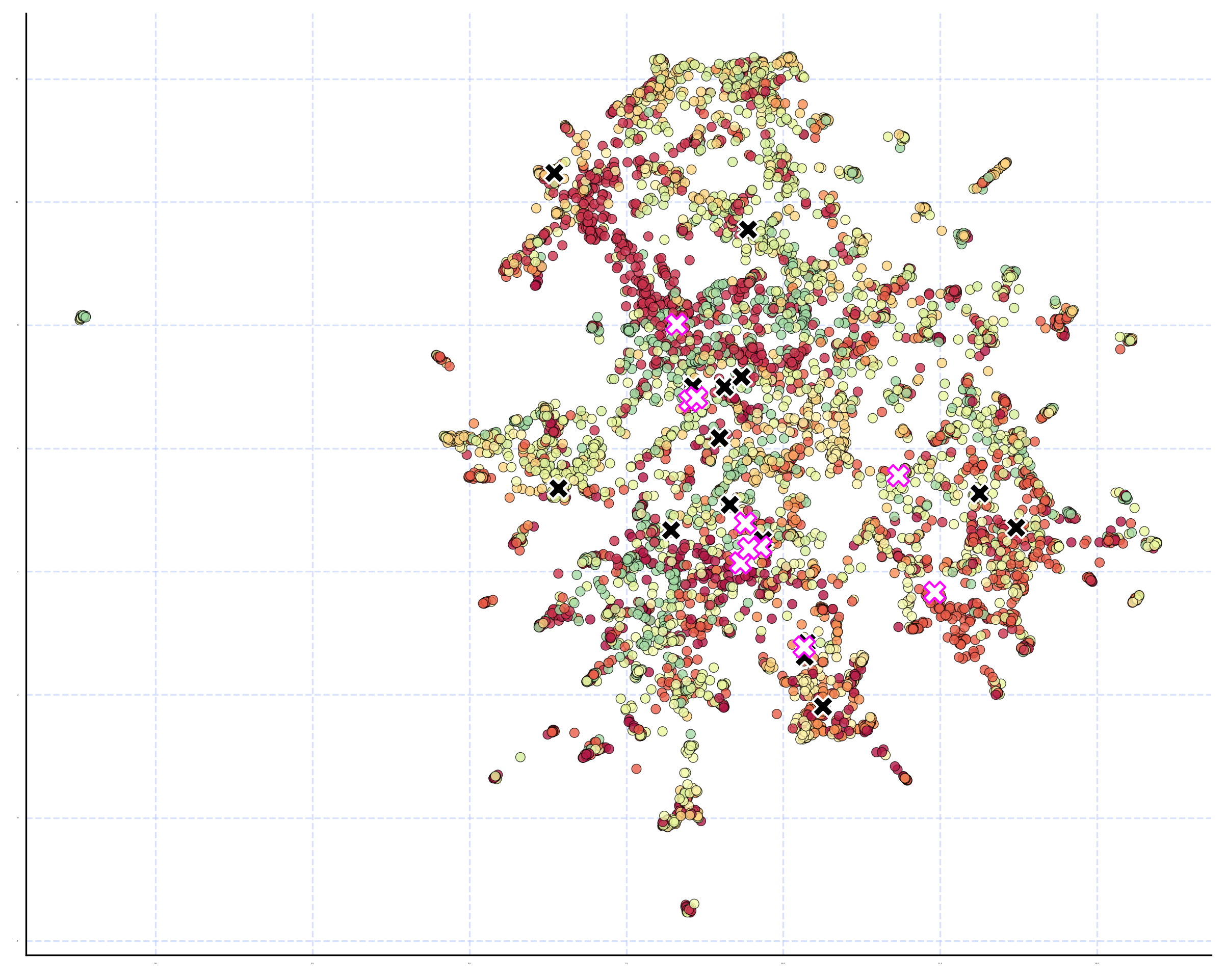}
    \caption{Quantum-refined clustering}
    \label{fig:run1_refined}
\end{subfigure}
\caption{Experiment 1 (k-Medoids, $k=10$) clustering visualization using 2D UMAP projection. Left: initial clustering with centroids (black), right: quantum-refined medoid selection with final centroids (pink).}
\label{fig:run1_clustering}
\end{figure}

Among all experiments, however, experiment 13 (GMM, k=10) achieved the highest nDCG (0.60) on training data, followed closely by experiment 17 (HDBSCAN-GMM, k=25) with 0.54, and Experiment 7 (HDBSCAN, k=10, no UMAP) with 0.52 and the lowest DBI overall (3.19). Experiment 13’s strong performance likely stemmed from GMM’s probabilistic flexibility at low k, which captured nuanced topical overlap and yielded the best retrieval quality. Experiment 17 benefited from HDBSCAN’s structure-aware initialization, followed by GMM fitting. This hybrid approach, especially at k=25, struck a strong balance between granularity and semantic coherence. In contrast, experiment 7 benefited from density-based clustering (HDBSCAN) applied directly to the high-dimensional space. At k=10, it effectively discovered dense semantic regions, while the quantum refinement stage helped consolidate them into meaningful, noise-filtered clusters.

\begin{figure}[H]
\centering
{\large \textbf{GMM Clustering Results (Experiment 13)}}
\vspace{0.3cm}

\begin{subfigure}{0.48\textwidth}
    \centering
    \includegraphics[width=\textwidth]{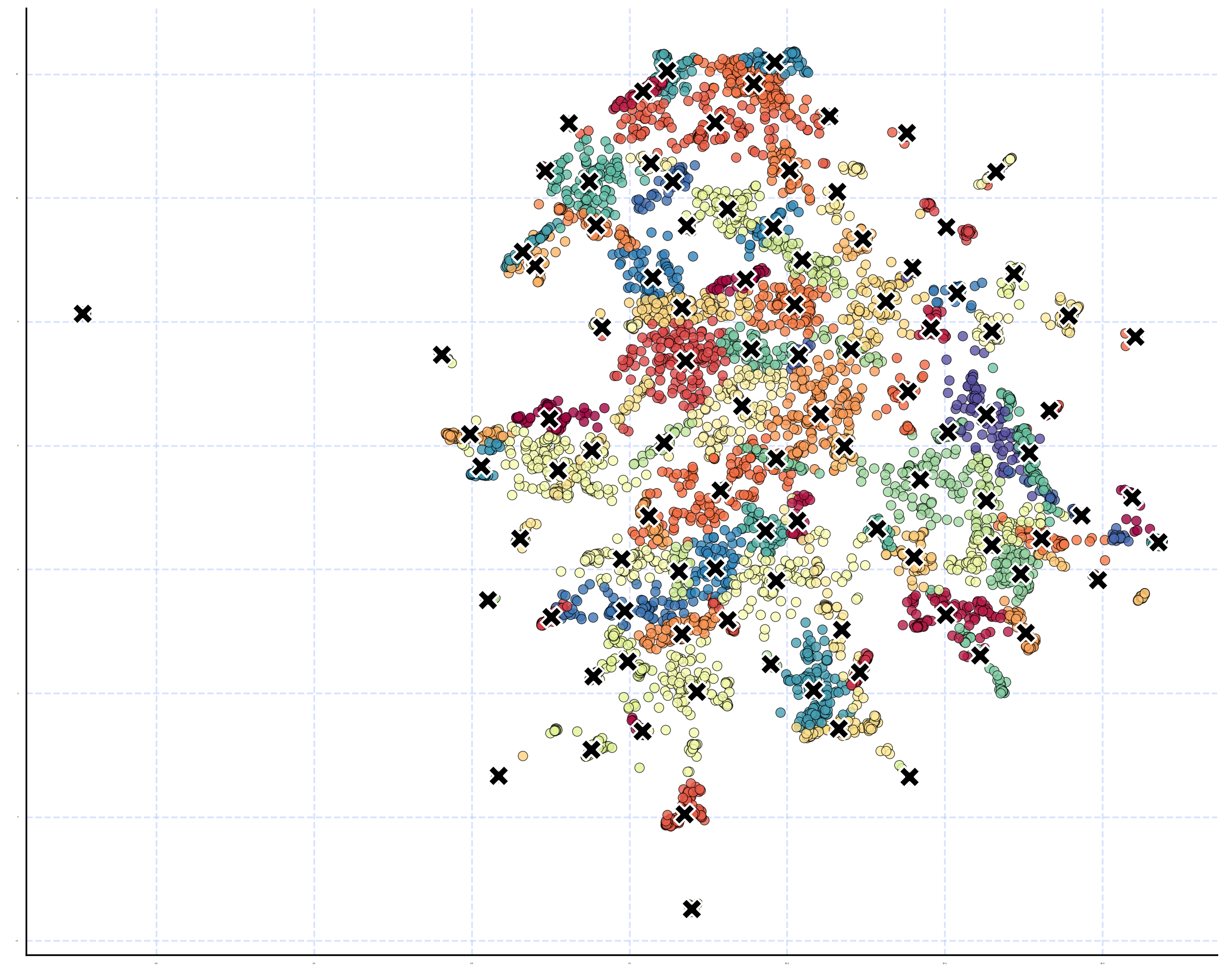}
    \caption{Initial clustering}
    \label{fig:run13_initial}
\end{subfigure}
\hfill
\begin{subfigure}{0.48\textwidth}
    \centering
    \includegraphics[width=\textwidth]{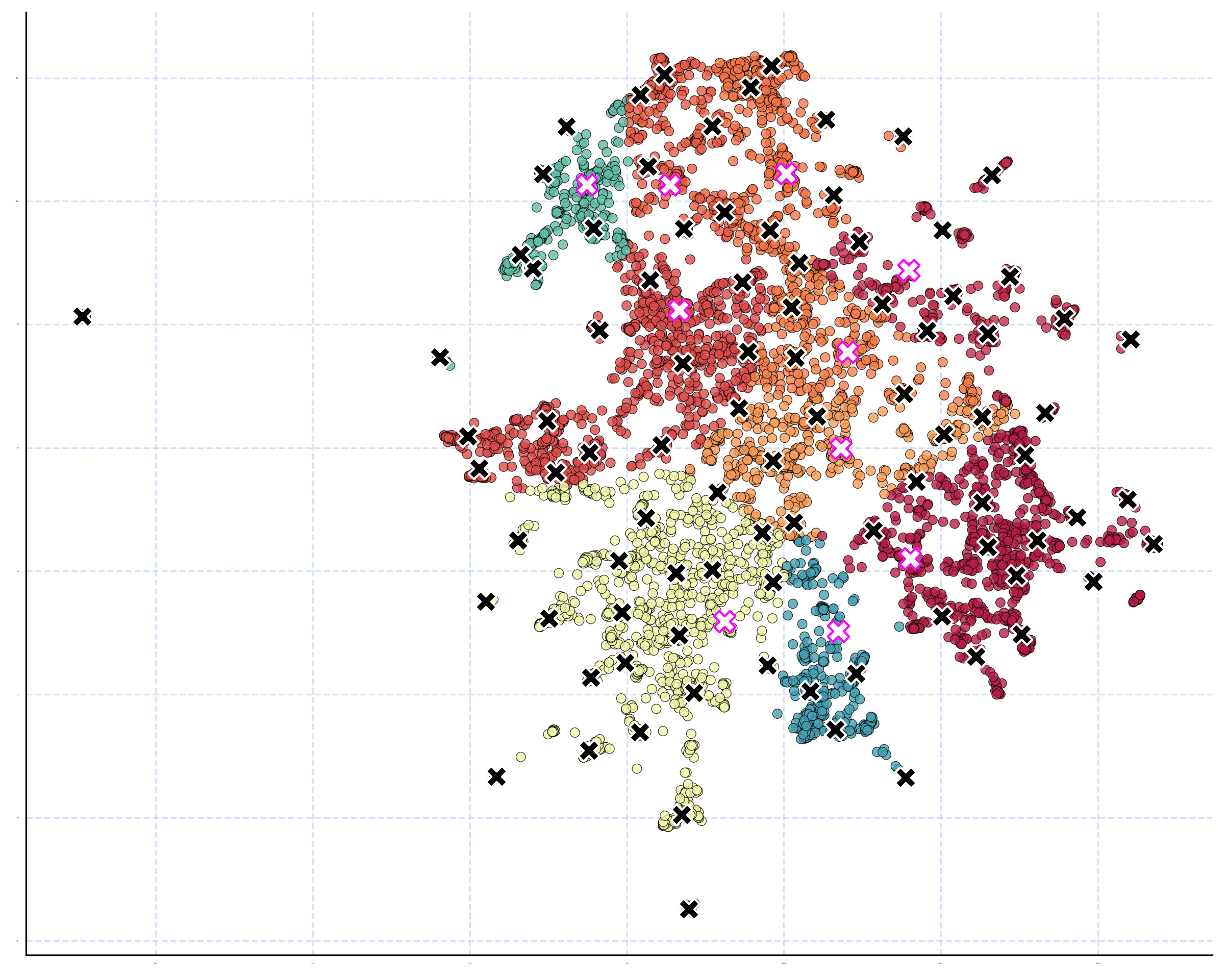}
    \caption{Quantum-refined clustering}
    \label{fig:run13_refined}
\end{subfigure}
\caption{Experiment 13 (GMM, $k=10$) clustering visualization using 2D UMAP projection. Left: initial clustering with centroids (black), right: quantum-refined medoid selection with final centroids (pink).}
\label{fig:run13_clustering}
\end{figure}

The results reveal consistent and interpretable trends across different clustering configurations, particularly with respect to the number of clusters (k), the use of dimensionality reduction, and the behavior of classical clustering methods prior to quantum refinement. Across all methods, increasing k generally led to improved Davies-Bouldin Index (DBI) scores, indicating tighter and more distinct clusters. This was most evident in the k-Medoids experiments, where DBI steadily decreased from 7.48 at k=10 to 3.71 at k=50, reflecting improved intra-cluster compactness and inter-cluster separation. However, while increasing k improved DBI, retrieval quality often peaked at lower k values using soft or structure-aware clustering methods. Baseline retrieval scores followed a similar pattern, dropping from 0.55 at k=10 to 0.47 at k=50, further emphasizing that expressiveness, not just compactness, plays a key role in modeling topical overlap in retrieval settings.

Dimensionality reduction using UMAP was explored as an optional preprocessing step to accelerate clustering and support visualization. While UMAP occasionally led to lower DBI scores as seen in experiments 7 and 8, its impact was not uniformly positive. In many cases, UMAP had little effect on DBI or even slightly worsened it. Moreover, improvements in geometric compactness did not consistently translate into better retrieval performance. In some configurations, especially at lower k, applying UMAP prior to clustering led to lower nDCG values, suggesting that key structural cues for retrieval may be lost in the projection to a reduced space. All GMM-based methods were run with UMAP applied due to their computational cost in high-dimensional space; non-reduced probabilistic clustering was excluded for tractability reasons, though it remains an area for future exploration.

Experiments 2 and 12, which applied UMAP before clustering, mistakenly submitted dimensionally reduced centroids to the leaderboard evaluation. Because retrieval metrics were computed using full-dimensional query embeddings, this mismatch resulted in invalid similarity calculations and artificially low leaderboard scores, especially for nDCG. These values should not be interpreted as indicators of poor clustering quality and instead reflect a representation mismatch during evaluation.

Together, these results validate the two-stage pipeline’s strategy of first generating an overcomplete and structurally diverse set of medoid candidates through classical clustering and then refining them using quantum-constrained optimization. The consistent improvements in DBI with higher k and the generally reliable performance of classical methods set a strong foundation for the second-stage quantum refinement, which enforces fixed-k constraints in a principled way.

\section{Future Work}

\subsection{Task 1: Feature Selection}

While our study focused on QUBO formulations built from combinations of mutual information, conditional mutual information, and permutation-based importance scores, there remain several promising directions for future exploration. We intend to extend our methodology to incorporate additional feature importance techniques inspired by classical literature. In particular, methods such as \textit{Functional ANOVA} (fANOVA)~\cite{fanova_hutter2014} and \textit{Leave-One-Feature-Out} (LOFO) importance~\cite{lofo_abid2020} offer intuitive measures of a feature's marginal and conditional relevance within a model context. These could potentially be adapted into the QUBO framework by mapping importance scores to diagonal entries and interactions (e.g., joint relevance or redundancy) to off-diagonal terms. Another promising candidate is the \textit{Relief} family of algorithms~\cite{relief_kira1992}, which estimate feature relevance based on how well feature values distinguish between near instances of different classes. Since Relief naturally accounts for both relevance and redundancy, it may be especially well-suited for QUBO-based optimization.

\subsection{Task 2: Instance Selection}

In the context of Task 2, it is important to note that all of the aforementioned computations were done on a per-batch basis. Thus every batch would have a most influential datapoint (with either svc or instance deletion based method) calculated only in relation to the other datapoints in the batch. We kept this as the same batching logic is applied to the off-diagonal terms. Two documents which are very highly similar (as measured by cosine similarity) to each other (and have either positive or negative label) can thus end up being in two different batches and their cosine similarity will be never taken into account. While batching is necessary as the QPU cannot fit all documents at once, this poses a challenge as the application of the ``penalties" and ``rewards" for hard instances are not applied on a global level. This is much easier to achieve at least for the diagonal elements however, as the influence points can be easily calculated only once and can be handled independently on the batch (e.g. an svc can be fit using all the training data and the distance from each instance to the support vector can be calculated for each instance, before the batching).

In addition, for the diagonal elements, more complex instance influence function methodology as in \cite{koh2017understanding},\cite{molnar2025}, \cite{joaquin2024in2core} can be applied either in combination with, or instead of the simple heuristics presented here.
 
\subsection{Task 3: Clustering}

In future work, we would like to evaluate how nDCG scores change when using the quantum processing unit. While we were unable to submit our most promising quantum experiments due to hardware and timing constraints, we remain curious how they would have scored under the competition’s official retrieval metrics on test data. Further future extensions could explore non-reduced probabilistic clustering to assess GMM performance in full embedding space. Additionally, incorporating probabilistic or fuzzy refinement in the quantum stage that may better capture semantic overlap in multi-topic documents.
\section{Conclusion}
In this work, we investigated the use of quantum annealing (QA) and simulated annealing (SA) to solve key machine learning optimization tasks - feature selection, instance selection, and clustering - by formulating them as QUBO problems. Across all tasks, we developed principled mappings that leveraged both classical and quantum resources effectively.

In Task 1 (feature selection), we explored multiple QUBO formulations that combined different feature importance and redundancy measures. Specifically, we tested combinations of MI, CMI, PFI, CPFI to construct the Q matrix. We evaluated these QUBOs using both simulated annealing and quantum annealing, and found that quantum annealing achieved comparable effectiveness to simulated annealing while requiring significantly less computational effort.

In Task 2 (instance selection), we extended the BCOS algorithm and introduced two new QUBO-based scoring mechanisms derived from SVM margins and instance deletion influence. Despite the lack of statisticaly significant differences between the methods at a reduction level of 25\%, these approaches showed promising results even at increased levels of instance reduction and can serve as a basis for further research on more difficult datasets. 

Task 3 (clustering) showcased the versatility of hybrid pipelines, combining classical clustering algorithms with quantum-constrained refinement. While the best overall clustering performance was achieved classically, our experiments confirmed that QUBO-based refinement enhances cluster diversity and compactness, particularly in document retrieval tasks.

Across tasks, we found that QA often matched or exceeded the performance of SA in less time, highlighting its potential for more efficient combinatorial optimization. While access to quantum annealers and scale remain ongoing challenges, our findings support the growing viability of quantum annealing as a practical tool in real-world ML pipelines.

\newpage
\begin{acknowledgments}
We thank the DS@GT CLEF team for providing valuable comments and suggestions. We would also like to thank Ayah Zaheraldeen and Jiangqin Ma for their input and support throughout the project. This research was supported in part through research cyberinfrastructure resources and services provided by the Partnership for an Advanced Computing Environment (PACE) at the Georgia Institute of Technology, Atlanta, Georgia, USA.
\end{acknowledgments}

\section*{Declaration on Generative AI}
 During the preparation of this work, the authors used OpenAI-GPT-4o: Grammar and spelling check. After using this tool, the authors reviewed and edited the content as needed and take full responsibility for the publication’s content.


\begin{thebibliography}{31}
\expandafter\ifx\csname natexlab\endcsname\relax\def\natexlab#1{#1}\fi
\providecommand{\url}[1]{\texttt{#1}}
\providecommand{\href}[2]{#2}
\providecommand{\path}[1]{#1}
\providecommand{\DOIprefix}{doi:}
\providecommand{\ArXivprefix}{arXiv:}
\providecommand{\URLprefix}{URL: }
\providecommand{\Pubmedprefix}{pmid:}
\providecommand{\doi}[1]{\href{http://dx.doi.org/#1}{\path{#1}}}
\providecommand{\Pubmed}[1]{\href{pmid:#1}{\path{#1}}}
\providecommand{\bibinfo}[2]{#2}
\ifx\xfnm\relax \def\xfnm[#1]{\unskip,\space#1}\fi
\bibitem[{Pasin et~al.(2025{\natexlab{a}})Pasin, Dacrema, Cuhna, Gon{\c{c}}alves, Cremonesi, and Ferro}]{qclef2025ceur}
\bibinfo{author}{A.~Pasin}, \bibinfo{author}{M.~F. Dacrema}, \bibinfo{author}{W.~Cuhna}, \bibinfo{author}{M.~A. Gon{\c{c}}alves}, \bibinfo{author}{P.~Cremonesi}, \bibinfo{author}{N.~Ferro},
\newblock \bibinfo{title}{Quantumclef 2025: Overview of the second quantum computing challenge for information retrieval and recommender systems at {CLEF}},
\newblock in: \bibinfo{editor}{G.~Faggioli}, \bibinfo{editor}{N.~Ferro}, \bibinfo{editor}{P.~Rosso}, \bibinfo{editor}{D.~Spina} (Eds.), \bibinfo{booktitle}{Working Notes of CLEF 2025 - Conference and Labs of the Evaluation Forum}, {CEUR} Workshop Proceedings, \bibinfo{year}{2025}{\natexlab{a}}.
\bibitem[{Pasin et~al.(2025{\natexlab{b}})Pasin, Dacrema, Cuhna, Gon{\c{c}}alves, Cremonesi, and Ferro}]{qclef2025lncs}
\bibinfo{author}{A.~Pasin}, \bibinfo{author}{M.~F. Dacrema}, \bibinfo{author}{W.~Cuhna}, \bibinfo{author}{M.~A. Gon{\c{c}}alves}, \bibinfo{author}{P.~Cremonesi}, \bibinfo{author}{N.~Ferro},
\newblock \bibinfo{title}{Overview of quantumclef 2025: The second quantum computing challenge for information retrieval and recommender systems at {CLEF}},
\newblock in: \bibinfo{editor}{J.~Carrillo{-}de{-}Albornoz}, \bibinfo{editor}{J.~Gonzalo}, \bibinfo{editor}{L.~Plaza}, \bibinfo{editor}{A.~G.~S. de~Herrera}, \bibinfo{editor}{J.~Mothe}, \bibinfo{editor}{F.~Piroi}, \bibinfo{editor}{P.~Rosso}, \bibinfo{editor}{D.~Spina}, \bibinfo{editor}{G.~Faggioli}, \bibinfo{editor}{N.~Ferro} (Eds.), \bibinfo{booktitle}{Experimental {IR} Meets Multilinguality, Multimodality, and Interaction. Proceedings of the Sixteenth International Conference of the {CLEF} Association (CLEF 2025)}, Lecture Notes in Computer Science, \bibinfo{year}{2025}{\natexlab{b}}.
\bibitem[{Farhi et~al.(2000)Farhi, Goldstone, Gutmann, and Sipser}]{farhi2000quantum}
\bibinfo{author}{E.~Farhi}, \bibinfo{author}{J.~Goldstone}, \bibinfo{author}{S.~Gutmann}, \bibinfo{author}{M.~Sipser},
\newblock \bibinfo{title}{Quantum computation by adiabatic evolution},
\newblock \bibinfo{journal}{arXiv preprint quant-ph/0001106}  (\bibinfo{year}{2000}).
\bibitem[{Boothby et~al.(2020)Boothby, Bunyk, Raymond, and Roy}]{boothby2020next}
\bibinfo{author}{K.~Boothby}, \bibinfo{author}{P.~Bunyk}, \bibinfo{author}{J.~Raymond}, \bibinfo{author}{A.~Roy},
\newblock \bibinfo{title}{Next-generation topology of d-wave quantum processors},
\newblock \bibinfo{journal}{arXiv preprint arXiv:2003.00133}  (\bibinfo{year}{2020}).
\bibitem[{Yulianti and Surendro(2023)}]{yulianti2023systematic}
\bibinfo{author}{L.~P. Yulianti}, \bibinfo{author}{K.~Surendro},
\newblock \bibinfo{title}{Implementation of quantum annealing: A systematic review},
\newblock \bibinfo{journal}{IEEE Transactions on Emerging Topics in Computing} \bibinfo{volume}{11} (\bibinfo{year}{2023}) \bibinfo{pages}{150--162}.
\bibitem[{Lucas(2014)}]{lucas2014ising}
\bibinfo{author}{A.~Lucas},
\newblock \bibinfo{title}{Ising formulations of many np problems},
\newblock \bibinfo{journal}{Frontiers in Physics} \bibinfo{volume}{2} (\bibinfo{year}{2014}) \bibinfo{pages}{5}.
\bibitem[{Date et~al.(2023)Date, Arthur, and Pusey‑Nazzaro}]{date2023qubo}
\bibinfo{author}{P.~Date}, \bibinfo{author}{D.~Arthur}, \bibinfo{author}{L.~Pusey‑Nazzaro},
\newblock \bibinfo{title}{Qubo formulations for training machine learning models},
\newblock \bibinfo{journal}{Quantum Computing Applications} \bibinfo{volume}{1} (\bibinfo{year}{2023}) \bibinfo{pages}{100--112}.
\bibitem[{M{\"u}cke et~al.(2023)M{\"u}cke, Heese, M{\"u}ller, Wolter, and Piatkowski}]{muecke2023feature}
\bibinfo{author}{S.~M{\"u}cke}, \bibinfo{author}{R.~Heese}, \bibinfo{author}{S.~M{\"u}ller}, \bibinfo{author}{M.~Wolter}, \bibinfo{author}{N.~Piatkowski},
\newblock \bibinfo{title}{Feature selection on quantum computers},
\newblock \bibinfo{journal}{Quantum Machine Intelligence} \bibinfo{volume}{5} (\bibinfo{year}{2023}) \bibinfo{pages}{11}. \URLprefix \url{https://doi.org/10.1007/s42484-023-00099-z}. \DOIprefix\doi{10.1007/s42484-023-00099-z}.
\bibitem[{Pranjić et~al.(2023)Pranjić, Mummaneni, and Tutschku}]{pranjic2023quantum}
\bibinfo{author}{D.~Pranjić}, \bibinfo{author}{B.~C. Mummaneni}, \bibinfo{author}{C.~Tutschku},
\newblock \bibinfo{title}{Quantum annealing based feature selection in machine learning},
\newblock \bibinfo{journal}{Quantum Machine Learning} \bibinfo{volume}{2} (\bibinfo{year}{2023}) \bibinfo{pages}{11--19}.
\bibitem[{Nembrini et~al.(2024)Nembrini, Dacrema, and Cremonesi}]{nembrini2023recommender}
\bibinfo{author}{R.~Nembrini}, \bibinfo{author}{M.~F. Dacrema}, \bibinfo{author}{P.~Cremonesi},
\newblock \bibinfo{title}{Feature selection for recommender systems with quantum computing},
\newblock \bibinfo{journal}{Journal of Computing Frontiers} \bibinfo{volume}{10} (\bibinfo{year}{2024}) \bibinfo{pages}{45--57}.
\bibitem[{Borle et~al.(2023)Borle, Zecevic et~al.}]{borle2023feature}
\bibinfo{author}{N.~Borle}, \bibinfo{author}{N.~Zecevic}, et~al.,
\newblock \bibinfo{title}{Feature selection with quantum annealing for interpretable and robust machine learning},
\newblock \bibinfo{journal}{Quantum Machine Intelligence} \bibinfo{volume}{5} (\bibinfo{year}{2023}) \bibinfo{pages}{1--15}.
\bibitem[{Pasin et~al.(2024)Pasin, Cunha, Gon{\c{c}}alves, and Ferro}]{pasin2024quantum}
\bibinfo{author}{A.~Pasin}, \bibinfo{author}{W.~Cunha}, \bibinfo{author}{M.~A. Gon{\c{c}}alves}, \bibinfo{author}{N.~Ferro},
\newblock \bibinfo{title}{A quantum annealing instance selection approach for efficient and effective transformer fine-tuning},
\newblock in: \bibinfo{booktitle}{Proceedings of the 2024 ACM SIGIR International Conference on Theory of Information Retrieval}, \bibinfo{year}{2024}, pp. \bibinfo{pages}{205--214}.
\bibitem[{Cunha et~al.(2023)Cunha, Fran{\c{c}}a, Fonseca, Rocha, and Gon{\c{c}}alves}]{cunha2023effective}
\bibinfo{author}{W.~Cunha}, \bibinfo{author}{C.~Fran{\c{c}}a}, \bibinfo{author}{G.~Fonseca}, \bibinfo{author}{L.~Rocha}, \bibinfo{author}{M.~A. Gon{\c{c}}alves},
\newblock \bibinfo{title}{An effective, efficient, and scalable confidence-based instance selection framework for transformer-based text classification},
\newblock in: \bibinfo{booktitle}{Proceedings of the 46th International ACM SIGIR Conference on Research and Development in Information Retrieval}, \bibinfo{year}{2023}, pp. \bibinfo{pages}{665--674}.
\bibitem[{Koh and Liang(2017)}]{koh2017understanding}
\bibinfo{author}{P.~W. Koh}, \bibinfo{author}{P.~Liang},
\newblock \bibinfo{title}{Understanding black-box predictions via influence functions},
\newblock in: \bibinfo{booktitle}{International conference on machine learning}, \bibinfo{organization}{PMLR}, \bibinfo{year}{2017}, pp. \bibinfo{pages}{1885--1894}.
\bibitem[{Molnar(2025)}]{molnar2025}
\bibinfo{author}{C.~Molnar}, \bibinfo{title}{Interpretable Machine Learning}, \bibinfo{edition}{3} ed., \bibinfo{year}{2025}. \URLprefix \url{https://christophm.github.io/interpretable-ml-book}.
\bibitem[{Joaquin et~al.(2024)Joaquin, Wang, Liu, Asher, Lim, Muller, and Chen}]{joaquin2024in2core}
\bibinfo{author}{A.~S. Joaquin}, \bibinfo{author}{B.~Wang}, \bibinfo{author}{Z.~Liu}, \bibinfo{author}{N.~Asher}, \bibinfo{author}{B.~Lim}, \bibinfo{author}{P.~Muller}, \bibinfo{author}{N.~F. Chen},
\newblock \bibinfo{title}{In2core: Leveraging influence functions for coreset selection in instruction finetuning of large language models},
\newblock \bibinfo{journal}{arXiv preprint arXiv:2408.03560}  (\bibinfo{year}{2024}).
\bibitem[{Bauckhage et~al.(2019)Bauckhage, Piatkowski, Sifa, Hecker, and Wrobel}]{bauckhage2019qubo}
\bibinfo{author}{C.~Bauckhage}, \bibinfo{author}{N.~Piatkowski}, \bibinfo{author}{R.~Sifa}, \bibinfo{author}{D.~Hecker}, \bibinfo{author}{S.~Wrobel},
\newblock \bibinfo{title}{A qubo formulation of the k-medoids problem.},
\newblock in: \bibinfo{booktitle}{LWDA}, \bibinfo{year}{2019}, pp. \bibinfo{pages}{54--63}.
\bibitem[{Alvarez-Giron et~al.(2024)Alvarez-Giron, Téllez-Torres, Tovar-Cortes, and Gómez-Adorno}]{AlvarezGiron2024qCLEF}
\bibinfo{author}{W.~Alvarez-Giron}, \bibinfo{author}{J.~Téllez-Torres}, \bibinfo{author}{J.~Tovar-Cortes}, \bibinfo{author}{H.~Gómez-Adorno},
\newblock \bibinfo{title}{Team qiimas on task 2 - clustering: Quantum annealing for k-medoids optimization},
\newblock in: \bibinfo{booktitle}{Working Notes of CLEF 2024 - Conference and Labs of the Evaluation Forum}, \bibinfo{address}{Grenoble, France}, \bibinfo{year}{2024}. \URLprefix \url{https://bitbucket.org/eval-labs/qc24-qiimas/src/main/}, \bibinfo{note}{cEUR Workshop Proceedings, ISSN 1613-0073}.
\bibitem[{Kurihara et~al.(2009)Kurihara, Tanaka, and Miyashita}]{kurihara2009quantum}
\bibinfo{author}{K.~Kurihara}, \bibinfo{author}{S.~Tanaka}, \bibinfo{author}{S.~Miyashita},
\newblock \bibinfo{title}{Quantum annealing for clustering},
\newblock in: \bibinfo{booktitle}{Proceedings of the Twenty-Fifth Conference on Uncertainty in Artificial Intelligence (UAI)}, \bibinfo{organization}{AUAI Press}, \bibinfo{year}{2009}, pp. \bibinfo{pages}{317--324}.
\bibitem[{Zaech et~al.(2023)Zaech, Danelljan, Birdal, and Van~Gool}]{zaech2023probabilistic}
\bibinfo{author}{J.-N. Zaech}, \bibinfo{author}{M.~Danelljan}, \bibinfo{author}{T.~Birdal}, \bibinfo{author}{L.~Van~Gool},
\newblock \bibinfo{title}{Probabilistic sampling of balanced k-means using adiabatic quantum computing},
\newblock \bibinfo{journal}{arXiv preprint arXiv:2310.12153}  (\bibinfo{year}{2023}). \URLprefix \url{https://arxiv.org/abs/2310.12153}.
\bibitem[{Matsumoto et~al.(2022)Matsumoto, Hamakawa, Tatsumura, and Kudo}]{matsumoto2022distance}
\bibinfo{author}{N.~Matsumoto}, \bibinfo{author}{Y.~Hamakawa}, \bibinfo{author}{K.~Tatsumura}, \bibinfo{author}{K.~Kudo},
\newblock \bibinfo{title}{Distance-based clustering using qubo formulations},
\newblock \bibinfo{journal}{Scientific Reports} \bibinfo{volume}{12} (\bibinfo{year}{2022}) \bibinfo{pages}{2669}. \URLprefix \url{https://doi.org/10.1038/s41598-022-06559-z}. \DOIprefix\doi{10.1038/s41598-022-06559-z}.
\bibitem[{Inc.(2023)}]{ocean}
\bibinfo{author}{D.-W.~S. Inc.}, \bibinfo{title}{Ocean software documentation}, \bibinfo{year}{2023}. \URLprefix \url{https://docs.ocean.dwavesys.com/}.
\bibitem[{Morstyn(2022)}]{pegasus_topology}
\bibinfo{author}{T.~Morstyn},
\newblock \bibinfo{title}{Annealing-based quantum computing for combinatorial optimal power flow},
\newblock \bibinfo{journal}{IEEE Transactions on Smart Grid} \bibinfo{volume}{PP} (\bibinfo{year}{2022}) \bibinfo{pages}{1--1}. \DOIprefix\doi{10.1109/TSG.2022.3200590}.
\bibitem[{Cover and Thomas(2006)}]{cover2006elements}
\bibinfo{author}{T.~M. Cover}, \bibinfo{author}{J.~A. Thomas}, \bibinfo{title}{Elements of Information Theory}, \bibinfo{edition}{2nd} ed., \bibinfo{publisher}{Wiley-Interscience}, \bibinfo{year}{2006}.
\bibitem[{Breiman(2001)}]{breiman2001random}
\bibinfo{author}{L.~Breiman},
\newblock \bibinfo{title}{Random forests},
\newblock \bibinfo{journal}{Machine Learning} \bibinfo{volume}{45} (\bibinfo{year}{2001}) \bibinfo{pages}{5--32}.
\bibitem[{Debeer and Strobl(2020)}]{debeer2020conditional}
\bibinfo{author}{D.~Debeer}, \bibinfo{author}{C.~Strobl},
\newblock \bibinfo{title}{Conditional permutation importance revisited},
\newblock \bibinfo{journal}{BMC Bioinformatics} \bibinfo{volume}{21} (\bibinfo{year}{2020}) \bibinfo{pages}{1--19}.
\bibitem[{Liu et~al.(2007)Liu, Pokharel, and Principe}]{liu2007correntropy}
\bibinfo{author}{W.~Liu}, \bibinfo{author}{P.~P. Pokharel}, \bibinfo{author}{J.~C. Principe},
\newblock \bibinfo{title}{Correntropy: Properties and applications in non-gaussian signal processing},
\newblock \bibinfo{journal}{IEEE Transactions on Signal Processing} \bibinfo{volume}{55} (\bibinfo{year}{2007}) \bibinfo{pages}{5286--5298}. \DOIprefix\doi{10.1109/TSP.2007.898255}.
\bibitem[{Pasvolsky and Inc.(2019)}]{dwave_combinations}
\bibinfo{author}{J.~Pasvolsky}, \bibinfo{author}{D.-W.~S. Inc.}, \bibinfo{title}{dimod.generators.combinations --- constraint generator for fixed-$k$ selection}, \bibinfo{year}{2019}. \URLprefix \url{https://github.com/dwavesystems/dimod/blob/main/dimod/generators/constraints.py}, \bibinfo{note}{accessed: 2025-06-12}.
\bibitem[{Hutter et~al.(2014)Hutter, Hoos, and Leyton-Brown}]{fanova_hutter2014}
\bibinfo{author}{F.~Hutter}, \bibinfo{author}{H.~H. Hoos}, \bibinfo{author}{K.~Leyton-Brown},
\newblock \bibinfo{title}{Efficient functional anova: Insights into high-dimensional model performance},
\newblock in: \bibinfo{booktitle}{Proceedings of the 30th Conference on Uncertainty in Artificial Intelligence (UAI)}, \bibinfo{year}{2014}.
\bibitem[{Abid et~al.(2020)Abid, Kamel, and Zou}]{lofo_abid2020}
\bibinfo{author}{A.~Abid}, \bibinfo{author}{A.~Kamel}, \bibinfo{author}{J.~Zou}, \bibinfo{title}{Lofo importance: Leave one feature out based feature importance score}, \bibinfo{howpublished}{\url{https://github.com/aerdem4/lofo-importance}}, \bibinfo{year}{2020}.
\bibitem[{Kira and Rendell(1992)}]{relief_kira1992}
\bibinfo{author}{K.~Kira}, \bibinfo{author}{L.~A. Rendell},
\newblock \bibinfo{title}{The feature selection problem: Traditional methods and a new algorithm},
\newblock \bibinfo{journal}{AAAI}  (\bibinfo{year}{1992}) \bibinfo{pages}{129--134}.

\end{thebibliography}

 
\end{document}